\documentstyle[epsfig]{elsart}
\journal{Optics Communications}
\begin{document}
%\draft
%---------------------------------------------------------------------------------
%\preprint{HEP/123-qed}
\begin{frontmatter}
\title{Propagation of Raman-matched laser pulses through a Bose-Einstein condensate}
\author{\"Ozg\"ur E. M\"ustecapl{\i}o\u{g}lu and L. You}
\address{
School of Physics, Georgia Institute of Technology, Atlanta GA
30309-0432, USA}
%\date{\today}
%\maketitle
%--------------------------------------------------------------------------
\begin{abstract}
We investigate the role of non-uniform spatial density profiles of
trapped atomic Bose-Einstein condensates in the propagation of
Raman-matched laser pulses under conditions for electromagnetically
induced transparency (EIT). We find that the sharp edged axial density
profile of an interacting condensate (due to a balance between
external trap and repulsive atomic interaction) is advantageous
for obtaining ultra slow averaged group velocities. Our results are
in good quantitative agreement with a recent
experiment report [Nature {\bf 397}, 594 (1999)].
\end{abstract}


\end{frontmatter}
%\narrowtext
%------------------------------------------------------------------------

\section{Introduction}
Recent reports \cite{haunew,ron} of stopping a near resonant laser pulse
inside highly absorptive atomic vapors has attracted tremendous public
interests. Physically, this magic like `light trapping' is due to
delicate quantum interference under the condition of
electromagnetically induced transparency (EIT)
\cite{eit,scully,fleischhauer-lukin}. In these experiments,
a dynamical controlled pump field effects the conversion of
a Raman-matched probe field into atomic Raman coherence of the medium.
Broadly interpreted, it provides a mechanism for storage of quantum
information encoded in the probe field. The experimenters also
demonstrated the read-out by back-converting stored atomic
Raman coherence into a new light field. EIT typically occurs
under conditions of complete opaqueness which prevents the
propagation of incoherent light; while a coherent
propagation can succeed because of interference induced
cancellation of one-path absorption.
In several earlier experiments, ultra slow light propagation was
observed in Bose-Einstein condensate (BEC) of Na atoms
[$\sim 17$ (m/s)\cite{hau} and $\sim 1$ (m/s)\cite{inouye}],
in hot rubidium vapor [$\sim 90$ m/s \cite{kash} and $\sim 8$ m/s
\cite{budker}], and in a Pr:YSO crystal [$\sim 45$ m/s
\cite{turukhin}]. As the net reduction of light speed
is about the same order of magnitude in different host
media, one may argue that being bose condensed is perhaps not so
crucial. From a practical point of view, however, the long
coherence time of BEC does prove to be advantageous.
Compared with a homogeneous media \cite{ron,kash,budker,turukhin},
using a trapped atomic BEC as the EIT medium
raises an important complication from its spatially inhomogeneous
density profile (due to both external trapping and
atom-atom interaction) \cite{haunew,hau,inouye}.
A consistent interpretation therefore requires careful
spatial averaging and theoretical modeling.
In this article we elucidate the effect of such a spatial average
which also takes into account atom-atom interaction.

Slow light propagation can be phenomenologically understood in terms
of a high index of refraction (slow phase velocity) or a high
dispersion-like feature (slow group velocity). It has been
extensively studied for atomic media exhibit EIT \cite{eit}.
Under typical conditions, such media are highly sought-after as
there exist many potential technological applications, e.g. optical delay
lines \cite{hau}, quantum entanglement of slow photons \cite{lukin-imamoglu},
non-classical (e.g. squeezed) and entangled atomic
ensembles \cite{lukin-yelin}, and quantum
memories \cite{haunew,ron,fleischhauer-lukin}.
Other potential fundamental applications include:
high nonlinear coupling between weak fields \cite{hau,harris},
quantum non-demolishing measurements and high
precision spectroscopy \cite{matsko}, and as
narrow-band sources for non-classical radiation
\cite{fleischhauer-lukin-matsko}. In addition to a complete
stoppage of light \cite{fleischhauer-lukin}, possibilities of
even negative group velocities have been discussed \cite{kocharovskaya}.
We expect our investigation will shed light onto the realization
of these applications in addition to providing a satisfactory
understanding of slow light propagation in trapped atomic
BEC.

The paper is organized as follows: In Sec. II we briefly
review our formulation of near-resonant light propagation
in dispersive medium. Section III is devoted to the
discussion of laser pulse propagation under conditions of
EIT from a pair of Raman-matched pulses. In Sec. IV we discuss
the widely used two-component model for the density profile of
a trapped interacting BEC.
In Sec. V we present numerical studies and compare them with
the earlier experimental report of Hau {\it et al.} \cite{hau}.
Finally we conclude in Sec. VI.

%\label{sec:intro}
%-------------------------------------------------------------------------

%-----------------------------------------------------------------------
\section{Group velocity in a dispersive media}
%\label{sec:wave_equation}
%------------------------------------------------------------------------
Propagation of light in a non-magnetic, charge-free, dispersive
medium is governed by the second order wave equation
\begin{eqnarray}\label{wave-eq}
\nabla^2_T\vec{E}(\vec{r},t)+\frac{\partial^2\vec{E}(\vec{r},t)}{\partial
z^2}-\frac{1}{c^2}\frac{\partial^2\vec{E}(\vec{r},t)}{\partial
t^2}+\vec{\nabla}\left(\vec{\nabla}\cdot\vec{P}(\vec{r},t)
\right)=\frac{4\pi}{c^2}\frac{\partial^2\vec{P}(\vec{r},t)}{\partial
t^2},
\label{eq1}
\end{eqnarray}
where $\vec{E}$ is the coarse grained electric field of
the propagating light and $\vec{P}$ is the induced macroscopic
medium polarization. Limiting our discussion to a
medium where spatial changes of the dielectric function are
negligible within a wavelength of propagating light, the last
term in the lhs of Eq. (\ref{eq1}) can be neglected. Although
this is a standard approximation, we cautiously point out that
it may lead to significant errors near the boundary of an
interacting atomic BEC, where the gas density profile changes
rapidly \cite{stringari}. In our numerical simulation we also
neglected this term based on an additional argument.
For a transverse field propagating along z-axis,
the coupling of $\vec{\nabla}\cdot\vec{P}(\vec{r},t)$ is
mainly along the orthogonal direction
from the propagation direction.
Therefore we only expect slight modifications to field distribution
in the immediate neighborhood of the BEC boundary.
We reconstruct the behavior of injected pulse in this region
from an interpolation between its behaviors in and out of the condensate.
The propagation is always started away from the edge of the BEC and
the thermal component density contribution of trapped atomic gas in
fully included. In the paraxial approximation limit, we make
further simplification of Eq. (\ref{eq1}) by neglecting the
transverse Laplacian $\nabla_T^2\vec{E}(\vec{r},t)$. This latter
approximation is checked for self-consistency in comparison with
experiments performed with a trapped BEC.

\begin{figure}[t]
\centerline{\epsfig{file=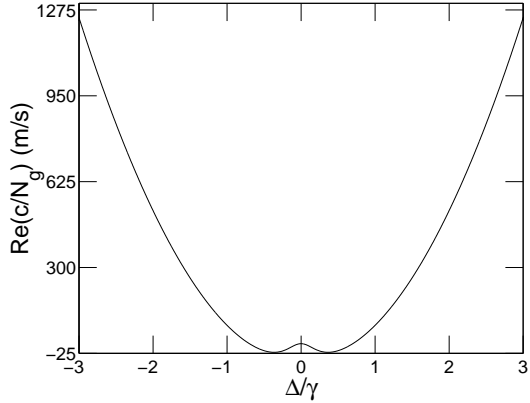,width=7.cm}\\[12pt]}
\caption{The group velocity for a pulse propagating through a
three-level Raman EIT medium of a uniform density calculated
according to $v_g=\Re(c/N_g)$. See later for the discussion gas
density $\sim 3.3\times10^{12}({\rm cm}^{-3})$,
$\lambda_0=589$(nm), $\Gamma_{31}=0.5\gamma$,
$\Gamma_{21}=(2\pi)10^3$ (Hz), $\Omega=0.56\gamma$, and
$\gamma=(2\pi)10.01$ (MHz).} \label{v1}
\end{figure}

\begin{figure}
\centerline{\epsfig{file=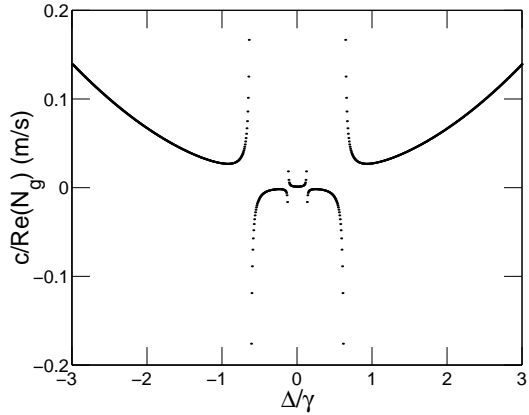,width=7.cm}\\[12pt]}
\caption{Same as in Fig. \ref{v1} except we used $v_g=c/\Re(N_g)$.
} \label{v2}
\end{figure}

We consider the propagation of a light pulse with central
wave vector $\vec{k}_0$ along the propagation ($z$) axis and
carrier frequency $\omega_0$. In a dispersive medium,
characterized by a complex refractive index $n(k,\omega)$ at
frequency $\omega$ and wave number $k$, the relation between the
carrier frequency and wavevector is the dispersion relation
$n(k_0,\omega_0)\omega_0=ck_0$. Typically $n(k_0,\omega_0)\approx 1$
at resonance, therefore $\omega_0=ck_0$
remains approximately valid throughout such a medium for
near-resonant carrier frequencies. Employing the slowly varying
phase and amplitude approximation for the complex amplitudes
$\{{\cal P},{\cal E}\}$, defined by $\{P,E\}=\Re[\{{\cal P},{\cal
E}\}\exp{(ik_0z-i\omega_0t)}]$, we obtain
\begin{eqnarray}\label{slwave-eq}
\frac{\partial{\cal E}}{\partial z}+\frac{1}{c}\frac{\partial{\cal
E}}{\partial t}=-\frac{4\pi}{c}\frac{\partial{\cal P}}{\partial
t}+2\pi i k_0{\cal P}.
\end{eqnarray}
The medium polarization is to be calculated from the linear
response of atoms to the weak probe field $E$.
It is customarily described in terms of an electric susceptibility
$\chi(\vec{r},t)$ according to
\begin{eqnarray}
{\cal P}(z,t)=\int\,dz^{\prime}\int\,dt^{\prime}
\chi(z-z^{\prime},t-t^{\prime}){\cal E}(z^{\prime},t^{\prime}).
\label{eq3}
\end{eqnarray}
All macroscopic physical quantities used are assumed to be
derived from a coarse graining procedure where the medium is assumed
isotropic microscopically. According to the convolution
theorem, the integral in the rhs of Eq. (\ref{eq3})
can be expressed as,
\begin{eqnarray}
{\cal P}(z,t)=\int\,d\omega \int\,d k \,\chi( k ,\omega){\cal E}(
k,\omega)e^{i kz-i\omega t},
\end{eqnarray}
which is equivalent to ${\cal P}(k,\omega)=\chi(k,\omega){\cal
E}(k,\omega)$. To simplify the notation, we will reference
$(k,\omega)$ with respect to the origin $(k_0,\omega_0)$.
Expanding $\chi(k,\omega)$ to first order
around $\chi_0=\chi(k_0,\omega_0)$ one obtains
\cite{harris-field-kasapi}
\begin{eqnarray}\label{pol-eq}
{\cal P}(z,t)=\chi(0){\cal
E}(z,t)+i\left(\frac{\partial\chi}{\partial\omega}\right)_0
\frac{\partial{\cal E}}{\partial t}
-i\left(\frac{\partial\chi}{\partial
k}\right)_0\frac{\partial{\cal E}}{\partial z},
\end{eqnarray}
which upon substituting into Eq. (\ref{slwave-eq})
yields \cite{kash,kocharovskaya}
\begin{eqnarray}
&&\left[1-2\pi k_0\left(\frac{\partial\chi}{\partial
k}\right)_0\right]\frac{\partial{\cal E}}{\partial z}+ \nonumber\\
&+& \frac{1}{c}\left[1+2\pi\chi(0)+2\pi \omega_0
\left(\frac{\partial\chi}{\partial\omega}\right)_0\right]\frac{\partial{\cal
E}}{\partial t} = 2\pi i k_0\chi(0){\cal E}. \label{e2}
\end{eqnarray}
Under the assumption of slow spatial variations we can
neglect the spatial dispersion term $k_0(\partial\chi/\partial k)_0$.
We now introduce a complex valued function
$N_g=1+2\pi\chi(0)+2\pi\omega_0\partial\chi/\partial\omega_0$,
whose real (imaginary) part will be called group index (phase correlation index).
The real (imaginary) part of $1+2\pi\chi(0)$ will be called
refractive (loss or gain) index. Unless otherwise stated,
we use $(.)^{\prime}$ and $(.)^{\prime\prime}$ to denote
real and imaginary part of (.) throughout this article.
The Wave equation (\ref{e2}) then simplifies to
\begin{eqnarray}\label{eq:wave}
\frac{\partial{\cal E}}{\partial
z}+\frac{1}{v_g}\frac{\partial{\cal E}}{\partial t}=\alpha{\cal
E},
\end{eqnarray}
where the complex parameter $\alpha=2\pi i k_0\chi(0)$
mainly governs pulse attenuation or amplification while the
complex velocity function $v_g=c/N_g$ is mostly responsible for
the propagation.
In order to appreciate the kinematic meaning of $v_g$, it is
helpful to cast the complex wave equation (\ref{eq:wave})
into two coupled equations for real functions  $\phi$ (phase)
and $U$ (amplitude) defined according to ${\cal E}=Ue^{i\phi}$,
\begin{eqnarray}
\frac{\partial U}{\partial z}+\frac{N_g^{\prime}}{c}\frac{\partial U}{\partial t}
&=&-2\pi k_0\chi^{\prime\prime}(0)U+\frac{N_g^{\prime\prime}}{c}
U\frac{\partial\phi}{\partial t}, \label{amp-eq}\\
\frac{\partial \phi}{\partial
z}+\frac{N_g^{\prime}}{c}\frac{\partial \phi}{\partial t} &=&2\pi
k_0\chi^{\prime}(0)-\frac{N_g^{\prime\prime}}{c}
\frac{\partial\ln{U}}{\partial t}. \label{phase-eq}
\end{eqnarray}
The propagation of a signal as well as the signal speed can only
be defined for such real observables.
When $N_g^{\prime\prime}\approx 0$, these two equations become
uncoupled with Eq. (\ref{amp-eq}) governing pulse propagation
kinematics. Consequently $v_g=c/N_g^{\prime}$ can be defined as
group velocity. Ambiguity arises when Eqs. (\ref{amp-eq}) and
(\ref{phase-eq}) are coupled for $N_g^{\prime\prime}\neq 0$.
Our choice of group velocity definition $v_g$ is
equivalent to $v_g=c/\Re(N_g)$, which is not the same as
$v_g=\Re(d\omega/dk)=\Re(c/N_g) $\cite{kocharovskaya}. When
$N_g^{\prime\prime}$ is not negligible these two definitions
generally lead to different results as can be exemplified by
selected results presented in Figs. \ref{v1} and \ref{v2},
which compares calculated group velocities according to
the above two different definitions.
The calculation of $N_g$ as well as the details of our model will
be given in the next section. At zero probe field detuning
$N_g^{\prime\prime}$ vanishes and the above two definitions
become identical. For detunings where $N_g^{\prime}$ vanishes,
however, the two results are drastically different as such points
are singularities for $v_g=c/N_g^{\prime}$. Thus, it seems more
appropriate to treat the group velocity as a complex function in
general, and assign a real physical meaning operationally based on
the pulse delays. The imaginary part $N_g^{''}$ causes non-trivial
coupling between the amplitude and phase of a propagating pulse.
As a result of this coupling, different propagation and
transmission characteristics can occur. This situation is similar
to the description of quantum tunneling in terms of a complex
time and velocity \cite{landauer}. When $N_g^{\prime\prime}$ is
large, the amplitude and phase equations are strongly coupled,
resulting in strong phase-amplitude correlations. This equation
set can also be compared to the semi-classical laser equation
\cite{scully} where similar situations arise. One important
consequence of such amplitude-phase correlations is a `false'
gain phenomenon, or a perceived superluminal propagation
\cite{super}. To illustrate such effects explicitly, we consider
a gas medium of uniform density. In this case, Eq.
(\ref{eq:wave}) can be solved with an ansatz
$\exp{[\alpha(z-z_0)]}f[t-(z-z_0)/v]$, where $z_0$ is the
injection location of the incoming pulse and $f(.)$ is an
arbitrary complex valued function at the complex retarded time
due to complex velocity $v=C/(N_g'+iN_g'')$. Assuming a Gaussian
temporal profile for the injected pulse at $z_0$ as
$\exp{(-at^2)}$, the phase and amplitude functions $U$ and $\phi$
can be solved
\begin{eqnarray}
U(z,t)&=&\exp{\left\{\alpha^{\prime}(z-z_0)+a\left[\frac{N_g^{\prime\prime}}{c}(z-z_0)\right]^2
-a\left[t-\frac{N_g^{\prime}}{c}(z-z_0)\right]^2\right\}},\\
\phi(z,t)&=&\alpha^{\prime\prime}(z-z_0)+2a\frac{N_g^{\prime\prime}}{c}(z-z_0)
\left[t-\frac{N_g^{\prime}}{c}(z-z_0)\right].
\end{eqnarray}
One can verify their correctness by direct
substitution into Eqs. (\ref{amp-eq}) and (\ref{phase-eq}).
We note that a non-vanishing $N_g^{\prime\prime}$ contributes to
$U$ as an amplification. Usually appreciable values of
$N_g^{\prime\prime}$ are found in the anomalous
dispersion regions where $N_g^{\prime}<0$.
A negative $N_g^{\prime}$ results in a negative delay time (or advance)
of pulse propagation, i.e. a superluminal propagation. Hence within
our general model, superluminal propagation is accompanied by
an amplification factor $\exp{[a(N_g^{\prime\prime}(z-z_0)/c)^2]}$
in addition to the nominal loss/gain factor of the
 medium $\exp{[\alpha^{\prime}(z-z_0)]}$.
This suggests that even in an absorptive medium with
$\alpha^{\prime}=-|\alpha^{\prime}|$, a superluminal pulse
can tunnel through provided that the medium produces strong
phase-amplitude correlation and is of a sufficient length.
Essentially one needs a medium of length
\begin{eqnarray}
L>\frac{|\alpha^{\prime}|c^2}{aN_g^{\prime\prime 2}},
\end{eqnarray}
to beat the absorption factor. To illustrate this point, we take
$\alpha^{\prime}=-0.1\,(1/\mu{\rm m})$, $N_g^{\prime}=-0.03c$,
$N_g^{\prime\prime}=0.03c$, typical numbers in the anomalous
dispersion region for near resonant pulse with a detuning
$\Delta\approx 0.2\gamma$, and $a=0.44\,(1/\mu{\rm s})^2$,
corresponding to a pulse of temporal width of $2.5\,(\mu{\rm s})$.
We obtain a critical value of $L\ge L_c=250\,(\mu{\rm m})$ for
superluminal propagation.

\begin{figure}
\centerline{\epsfig{file=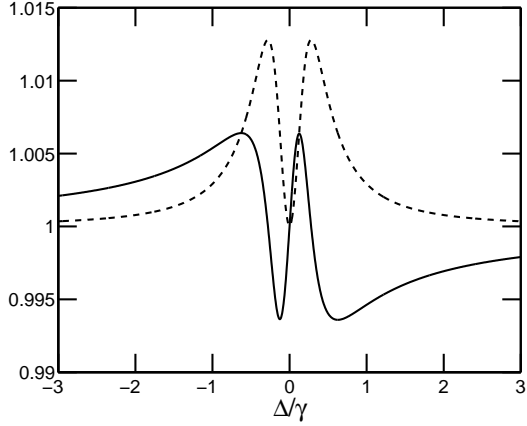,width=7.cm}\\[12pt]}
\caption{Refractive index (solid curve) and loss index (dashed
curve) for an atomic cloud with a density $3.3\times10^{12}({\rm
cm}^{-3})$ corresponding the peak density of a trapped BEC at
$T\approx 450$ (nK), slightly above $T_C$ [6]. Other parameters
used are: $\lambda_0=589$ (nm), $\Gamma_{31}=0.5\gamma$,
$\Gamma_{21}=(2\pi)10^3$ (Hz), $\Omega=0.56\gamma$,
$\gamma=(2\pi)10.01$ (MHz).} \label{fig3.eps}
\end{figure}

\begin{figure}
\centerline{\epsfig{file=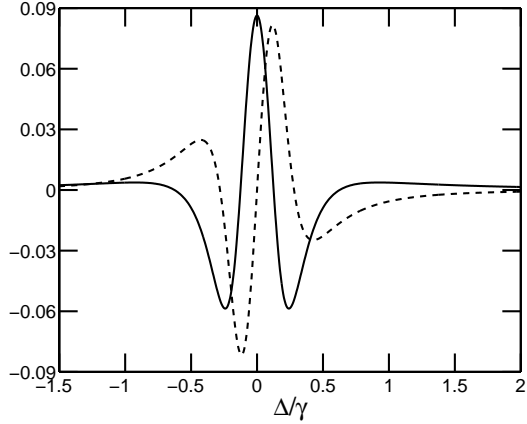,width=7.cm}\\[12pt]}
\caption{Group index (solid line) and phase correlation index
(dashed line) (scaled by $1/c$) under the same conditions as Fig.
\ref{fig3.eps}.} \label{fig4.eps}
\end{figure}

\begin{figure}[t]
\centerline{\epsfig{file=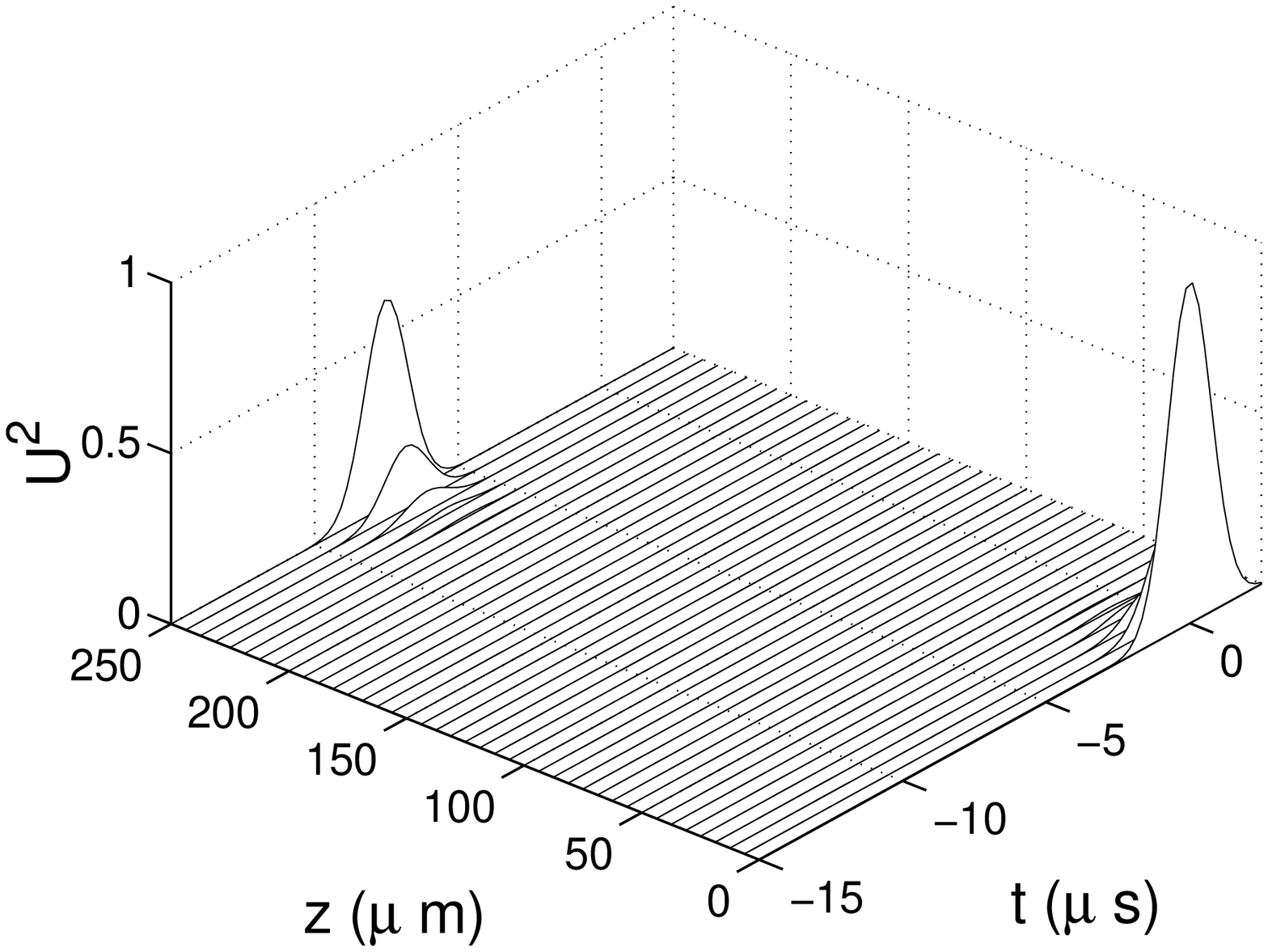,width=7.cm}\\[12pt]}
\caption{Superluminal pulse propagation, or
 {\it tunneling} through an EIT medium.
The negative delay time of $7.5 (\mu{\rm s})$ corresponds to a
velocity higher than $c$ by $\sim 30$ (m/s).} \label{fig5.eps}
\end{figure}

In Fig. \ref{fig4.eps}, we notice that $N_g^{\prime}$ has a local
minimum at $\Delta\approx 0.3\gamma$. This local minima takes a
negative value which signals a maximum negative delay time.
However, in this case $N_g^{\prime\prime}$ is vanishingly small
and it requires $L\rightarrow\infty$ to beat the absorption. With
careful analysis, one can choose an optimum value of
$N_g^{\prime}$ to take advantage of the amplification due to a
nonzero $N_g^{\prime\prime}$. A typical simulation for the
physical observable $U^2(z,t)$ is given in Fig. \ref{fig5.eps}.
For the parameters used, a temporal advancement of the pulse, in
other words a negative delay of $7.5 (\mu{\rm m})$ is found, which
corresponds to a velocity higher than $c$ by $\sim 30 ({\rm
m/s})$. It is possible to get higher superluminal velocities by
decreasing the atomic vapor density. For instance, at
$\rho=3.3\times10^{10} ({\rm cm}^{-3})$, which is $100$ times
smaller than used in Figs. \ref{fig3.eps} and \ref{fig5.eps}, the
relevant parameters become $N_g^{\prime}=-3\times10^{-4}c,
N_g^{\prime\prime}=3\times10^{-4}c$, and $\alpha^{\prime}=-1000
({\rm 1/cm})$. In this case, superluminal tunneling occurs for
$L>2 ({\rm cm})$. These parameters are experimentally accessible
and in fact are close to reported values in a recent experiment
\cite{akulshin}, where $N_g^{\prime}=-8\times10^{-5}c$ is measured
in a Rb vapor of length $L=5 ({\rm cm})$.

To develop an operational theory for the slow light propagation
as in the experiment \cite{hau}, we need to include the above
discussed amplitude-phase correlation as well as
the non-uniform atomic gas density distribution.
Furthermore, it would be necessary to include transverse
diffraction losses as they might contribute significantly
as the axial length of a trapped BEC increases.
With all these complications, we have to resort to numerical
simulations with available experimental parameters.
Thus, we cannot simply define a group velocity for a pulse
as $c/N_g^{\prime}$ or some other form. We should define a
viable speed in accordance with the actual propagation of
a physical pulse. In this article we follow the operational
definition as used in experiments \cite{hau,inouye}.
We will concentrate on ultraslow pulse propagation
rather than the equally likely superluminal effects.

The existence of amplitude-phase correlation calls for caution
in interpreting both slow and fast light propagation in a dispersive
or active medium. Straight forward conclusions based on a linear
response theory and simple definitions of group velocities
can not be taken very seriously, especially when conflicting
view with non-locality properties of electromagnetic fields
arise. Strictly speaking, a pulse is composed of Fourier
superposition
of infinite wave trains with potentially different phase velocities.
It is meaningful only when there exists a finite bandwidth
around central frequency so that we can write \cite{stratton}
\begin{eqnarray}
E(z,t)=\int_{k_0-\delta k}^{k_0+\delta k} E(k)e^{ikz-i\omega(k) t}
dk.
\end{eqnarray}
Within a small bandwidth, linearization of the frequency around
the center carrier gives $\omega(k)\approx\omega_0+v_g(k-k_0)$ with
$v_g=d\omega/dk$ (evaluated at $k_0$) the perceived group velocity.
Such a velocity represents the speed of propagation of
constant pulse amplitude surface, a non-existent situation when
strong phase-amplitude correlation exists. The large cross
phase-amplitude due to loss/gain index (or the imaginary part of the
group velocity) causes ambiguity since the constant amplitude
surface follows a trajectory now in the complex plane. As a result,
we recommend treating the group velocity as a complex valued function.
Further discussions will depend on specifics of
medium's electric susceptibility. This is the subject
of the next section.
%--------------------------------------------------------------------
\section{A Raman-matched EIT medium of cold atoms}
\label{sec:eit-bec}
%---------------------------------------------------------------------
Our study in this article is based on a medium of three-level
$\Lambda$-type atoms. The excited level $|3\rangle$ is assumed to
couple to the final state level $|2\rangle$ by a dressing laser beam
with a Rabi frequency $\Omega$, and to the ground state $|1\rangle$
by the probe beam, whose propagation is the central issue of study.
The dressing pump laser is assumed to be nearly Raman-matched with
the probe laser (of carrier frequency $\omega_0$), which is detuned
by $\Delta=\omega_0-\omega_{31}$ from the
$|3\rangle\rightarrow|1\rangle$ transition frequency
$\omega_{31}$. Initially all atoms are assumed to be in the
ground state level $|1\rangle$. In the weak probe field limit,
a linear response calculation can be carried as only a small fraction of
atoms are pumped out of their initial state. In this study we
use the approximation that the initial density profile $\rho_1(\vec{r})$
being unchanged. Solving the density matrix equation for atoms,
we determine the linear susceptibility to be $\chi=\eta\chi_1$,
directly related to  the steady state value
for the $|3\rangle\rightarrow|1\rangle$ matrix element. Here
$\eta(\vec{r})=(3\lambda_{31}^3/32\pi^3)\rho_1(\vec{r})$ and
$\lambda_{31}$ is the transition wavelength and \cite{eit}
\begin{eqnarray}
\chi_1(\Delta,\Omega)=-\left[\frac{\Delta}{\Gamma_{31}}+i\left(1+
\frac{\Omega^2}{4\Gamma_{31}(\Gamma_{21}-i\Delta)}\right)\right]^{-1}.
\end{eqnarray}
$\Gamma_{31}$ is the decay rate from $|3\rangle$ to $|1\rangle$
and $\Gamma_{21}$ is the decoherence rate between $|2\rangle$ and
$|1\rangle$. Typical detuning dependence of refractive and loss
indices calculated from this susceptibility are shown in Fig.
\ref{fig3.eps}. The twin peaks in the loss index are from dressed
absorption lines (Autler-Townes doublet \cite{dp}), and the narrow
valley between them is associated with the normal dispersion
region in the refractive index curve. Within the valley region,
absorptive loss is vanishingly small. Far away from this lossless
normal dispersion region both curves diminish resulting in a
marked increase of transmission, except around the absorption
peaks, where optical response is dominated by normal dispersion.
Anomalous dispersion regions are accompanied by significant
absorptive loss. For comparison, the group and phase correlation
indices are illustrated in Fig. \ref{fig4.eps}. In both normal and
anomalous dispersion regions, we note that $N_g^{\prime}$ and
$N_g^{\prime\prime}$ are now dominated by the term
($\partial\chi/\partial\Delta$) due to sharp variations in
refractive and loss indices. In the normal dispersion region
$N_g^{\prime}>0$, and particularly in the almost lossless region
it attains a maximum on resonance,. This produces a  minimal and
well defined group velocity as $N_g^{\prime\prime}\approx 0$ in
the same location. In the anomalous region $N_g^{\prime}$ becomes
negative or zero, corresponding to vanishingly small or negative
delay times of propagation, as recently observed in \cite{budker}.
The frequency derivative $\partial\chi/\partial\Delta$ is found to
be $\eta\partial\chi_1/\partial\Delta$ with
\begin{eqnarray}
\frac{\partial\chi_1}{\partial\Delta}=\frac{\frac{1}{\Gamma_{31}}
-\frac{\Omega^2}{4\Gamma_{31}(\Gamma_{21}-i\Delta)^2}}{\left[\frac{\Delta}{\Gamma_{31}}+i\left(1+
\frac{\Omega^2}{4\Gamma_{31}(\Gamma_{21}-i\Delta)}\right)\right]^2}.
\end{eqnarray}
We can then define an auxiliary function
$h(\Delta,\Omega)=\chi_1+(\Delta+\omega_{31})\partial\chi_1/\partial\Delta$,
to express $N_g=1+2\pi\eta h(\Delta,\Omega)$. We note that
$N_g^{\prime}$ is maximum on resonance, and the trapped atomic
density profile peaks at $\vec{r}=0$. Therefore an estimate of
the minimum for group speed is
$v_g^0=c/[1+2\pi\eta(0)h^{\prime}(0,\Omega)]$. For a sample size
of length $L$, we introduce dimensionless variables
$\bar{z}=2z/L$ and $\bar{t}=(v_g^0/L)t$, and a normalized density
profile function $f(\vec{r})=\rho_1(\vec{r})/\rho_1(0)$. In terms
of these parameters, the wave equation Eq. (\ref{slwave-eq})
becomes
\begin{eqnarray}
\frac{\partial{\cal E}}{\partial\bar{z}}+\left[\frac{v_g^0}{2c}+
\frac{h(\Delta,\Omega)}{2h^{\prime}(0,\Omega)}f(\vec{r})\right]
\frac{\partial{\cal E}}{\partial \bar{t}}=\frac{i\pi L
}{2\lambda_{31}}\eta(0)\chi_1(\Delta,\Omega)f(\vec{r}){\cal E}.
\label{seq}
\end{eqnarray}
In a typical experiment with $\Delta=0$ and $\Gamma_{31}\gg\Gamma_{21}$,
we find
\begin{eqnarray}
\chi_1(0,\Omega)&&\approx
i\frac{4\Gamma_{31}\Gamma_{21}}{\Omega^2},\nonumber\\
\left(\frac{\partial\chi_1}{\partial\Delta}\right)_{\Delta=0}&&\approx\frac{4\Gamma_{31}}{\Omega^2}.
\end{eqnarray}
Thus, $N_g^{\prime\prime}$ is indeed negligible and we
have $N_g\approx 2\pi\omega_{31}\eta(4\Gamma_{31}/\Omega^2)$.
In this case, a well defined group velocity exists and is given by
\begin{eqnarray}
v_g=c/N_g^{\prime}\approx \frac{4\pi^2\Omega^2c}{\rho
3\lambda_{31}^3\Gamma_{31}}\omega_a=\frac{\hbar
c\epsilon_0\Omega^2}{2\omega_{31}d_{31}^2\rho}.
\end{eqnarray}
For convenience, the last expression is in SI units
($d_{31}^2/\hbar=3\lambda_{31}^3\Gamma_{31}/32\pi^3$).
This result agrees with the one used for estimations of
experimental results \cite{hau,morigi}. It is also the same
as obtained earlier \cite{harris,grobe,xiao,kasapi}.
When applied to trapped atomic BEC at ultra low temperatures,
inhomogeneous density profile needs to be included properly.
When local density approximation is assumed, we expect the local light
velocity should display rapid slowing
down upon entering the BEC region from the thermal component and
rapid acceleration when exiting it. This in general will result in
different averaged velocity from simple estimations based on
a uniform density. In order to elucidate this
effect, it is necessary to specify an explicit form for the
density profile of the BEC.
%----------------------------------------------------------------------
\section{Spatial density profile of an interacting BEC}
%----------------------------------------------------------------------
In this section, we review briefly a simple analytical model
for the density profile of an interacting BEC as developed
in Ref. \cite{naraschewski}.
We take the bose gas (below $T_c$) as composed of two components;
a condensate part, whose density is computed using the
Thomas-Fermi approximation and a background thermal atom part
whose density is computed semi-classically assuming a continuum
density of states in a harmonic trap \cite{naraschewski,bagnato}.
The total ground state density is then found to be
\begin{eqnarray}\label{bec}
  \rho_1(\vec{r}) =
  \frac{\mu-V}{U}
  \theta(\mu-V)\theta(T_C-T)
  +\frac{g_{3/2}({\sf z} e^{-\beta V})}
  {\Lambda_T^3},
\end{eqnarray}
where $U=4\pi\hbar^2 a_{\rm sc}/M$. $a_{\rm sc}$ is the atomic
scattering length and $\theta(.)$ is the Heaviside step function,
$g_{n}(x)=\sum_j x^j/j^{n}$, $\Lambda_T$ is the thermal de
Br\"oglie wavelength and $\beta=1/(k_BT)$. The external trapping
potential is $V(\vec{r})=M\omega_r^2(r^2+\epsilon^2 z^2)/2$ with
$\omega_r$ the radial trap frequency and $\epsilon$ the aspect
ratio. $M$ is atomic mass. The chemical potential $\mu$
is determined by the normalization $N=\int
d\vec{r}\rho_F(\vec{r})$ with $N$ the total number of atoms.
At high temperatures it is determined by solving ${\rm Li}_3({\sf
z})=(T/T_C)^{-3}\zeta(3)$ for fugacity ${\sf z}=e^{\beta\mu}$.
${\rm Li}_3(.)$ and $\zeta(3)$ are the third order polylogarithm
and Riemann-Zeta functions respectively. At low temperatures, it
is found that \cite{naraschewski},
\begin{eqnarray}
  \mu=\mu_{TF} \left( \frac{N_0}{N} \right)^{2/5},
  \label{mu-bec}
\end{eqnarray}
with $\mu_{TF}$ the Thomas-Fermi approximation to chemical
potential and the condensate fraction
\begin{eqnarray}
 \frac{N_0}{N} =
 1-\left(\frac{T}{T_C}\right)^3-s\frac{\zeta(2)}{\zeta(3)}\left(\frac{T}{T_C}
 \right)^2\left[1-\left(\frac{T}{T_C}\right)^3\right]^{2/5},
 \label{cond-frac}
\end{eqnarray}
with a scaling parameter $s$,
\begin{eqnarray}
  s =
  \frac{\mu_{TF}}{k_BT_C}=\frac{1}{2}\zeta(3)^{1/3}\left(15N^{1/6}\frac{a_{\rm sc}}
  {a_{ho}}\right)^{2/5}.
  \label{scaling}
\end{eqnarray}
$a_{\rm ho}=\sqrt{\hbar/(M\epsilon^{1/3}\omega_r)}$ denotes
the average harmonic oscillator length scale \cite{giorgini}.
%-----------------------------------------------------------------------------------
\section{Numerical results and discussion}
\label{sec:numerics}
In this section, we discuss several aspects of our numerical
investigation for the experiment \cite{hau}.
%-------------------------------------------------------------------------------------
\subsection{Group velocity in ideal and
interacting BEC}
\begin{figure}[t]
\centerline{\epsfig{file=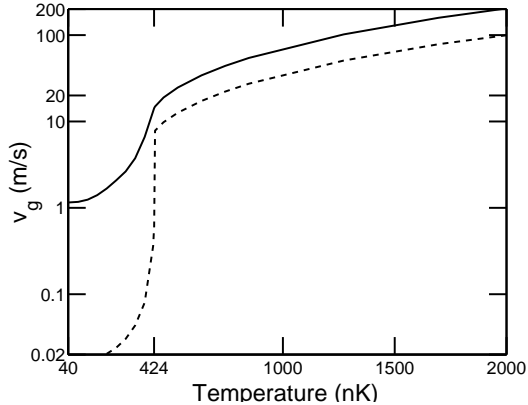,width=7.cm}\\[12pt]}
\caption{Comparison of an interacting (solid line) and ideal
(dashed line) models for the temperature dependence of
operationally defined group velocity. We have used the same
parameters as in in Fig. \ref{fig3.eps}. In addition,
$N=8.3\times10^6$, $a_s=2.75$ (nm), $\omega_r=(2\pi)69$ (Hz),
$\omega_z=(2\pi)21$(Hz), $M=23$ (amu) as for the experiment [6].
$R_0=15$ ($\mu$m).} \label{fig6}
\end{figure}

\begin{figure}
\centerline{\epsfig{file=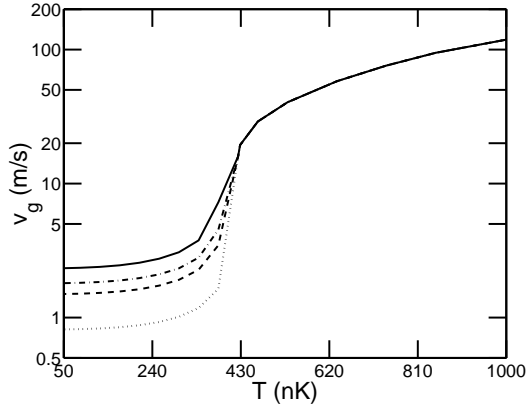,width=7.cm}\\[12pt]}
\caption{Group velocity is calculated with interacting BEC for
$a_{\rm sc} = 7,5.75,3.75,1$(nm) corresponding to solid,
dot-dashed, dashed, and dotted curves, respectively. Other
parameters are the same with those in Fig. \ref{fig6}. }
\label{vg-as}
\end{figure}

According to experimental prescriptions \cite{hau,inouye}
we take the operational approach, first determine the delay
time according to
\begin{equation}\label{group-delay}
t_d(T) = \frac{1}{\pi R^2}\int_0^R 2\pi dr
\int_{-L}^{L}dz\frac{1}{v_g(\vec{r})}-\frac{2L}{c},
\end{equation}
where $R$ is the pinhole radius introduced to selectively detect
propagated light from the near axial region of the atomic cloud.
As justified before \cite{ozgur1} we will ignore the spatial $(R)$
dependence of $L$ due to the shape of the cloud for small pinhole
sizes and relatively long optical path length $L$ \cite{morigi}.
Within the local density approximation, the local group velocity
is given by $v_g(\vec{r})=c/N_g^{\prime}$, and its average is then
calculated according to $\langle v_g\rangle = L(T)/t_d(T)$ with
$L(T)$ is the on-axis cloud size at a given temperature $T$. In
Fig. \ref{fig6} we compare the calculated results for an
interacting and an ideal atomic gas.
For the ideal gas case, our results simply reproduced those in
Ref. \cite{morigi}. Whenever possible we have used parameter values
taken from Hau's experiment \cite{hau}. At high temperatures both
models are in agreement with the experimental data. At low
temperatures, our model including atom-atom interaction gives
results of the correct order of magnitude, while those
based on ideal non-interacting atoms \cite{morigi} are two orders
of magnitude too low. We also note that the sharp low temperature
dependence near the BEC transition temperature $T_C$ becomes
significantly smoother, i.e. closer to the experimental
observations when the atom interaction is included. Subsequently
the calculated $\langle v_g\rangle$ does not fall down as rapidly
at lower temperatures. Our model including atom-atom interaction
also predicts that $\langle v_g\rangle$ increases with the atomic
scattering length $a_{\rm sc}$ as illustrate in Fig. \ref{vg-as},
where the group velocity is calculated for various scattering length
parameters $a_{\rm sc} = 7,5.75,3.75,1\,$(nm), corresponding to $s
= 0.3982,0.368,0.31,0.183$, respectively. We note that the two
component density profile works particularly well for $s\ll 1$,
but becomes less accurate when $s>0.3$ \cite{naraschewski}. Thus,
we conclude that the inclusion of interatomic interactions improves
upon the ideal gas model prediction both qualitatively and
quantitatively. In fact, our results with atom interaction give
the correct order of magnitude as reported in the experiment \cite{hau}.
Although we are still not near the reported value
$v\sim 17\,({\rm m/s})$ for $T\sim50\,({\rm nK})$. Inclusion of
condensate number fluctuations, evaporation dynamics, and a
self-consistent treatment of density profile might
bring in further improvements.

\subsection{Axial propagation of probe pulse}
\begin{figure}
\centerline{\epsfig{file=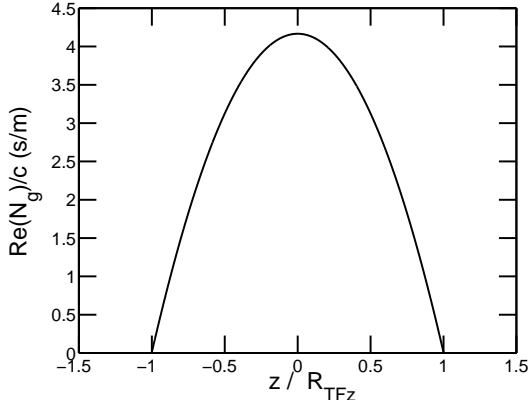,width=7.cm}\\[12pt]}
\caption{Axial profile of the scaled group index $N_g^{\prime}/c$
at $T=43$ (nK) and $\Delta=0$. $N_g^{\prime\prime}/c$ is very
small with a peak value $\sim10^{-12}$(s/m). All parameters are
the same as in Fig. \ref{fig6}.} \label{graded.eps}
\end{figure}

\begin{figure}[t]
\centerline{\epsfig{file=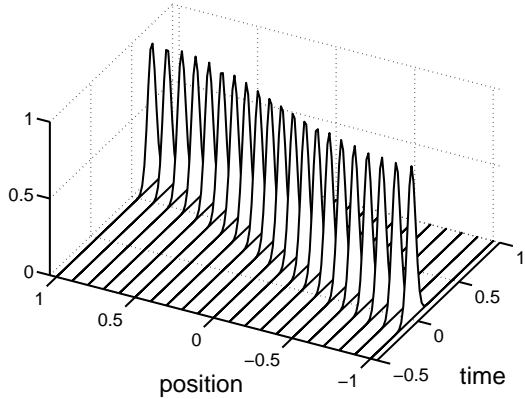,width=7.cm}\\[12pt]}
\caption{Propagation of a resonant pulse with $\Delta=0$, $T=43$
(nK). All parameters are the same as in Fig. \ref{fig6}. Position
and time are given in dimensionless units explained in the text.}
\label{fig9}
\end{figure}

At temperatures well below $T_C$, both axial and radial density
profiles of an atomic cloud are dominated by its condensate
component. For the harmonic trapping potentials considered here,
the Thomas-Fermi approximation results in an inverted paraboloid
density profile. Consequently, the spatial behavior of the susceptibility
in the linear response regime also follow such a behavior as shown
in Fig. \ref{graded.eps}.
A typical simulation for the time delay of a resonant pulse
$\exp{(-100 t^2)}$ after its passage through such a medium is
is shown in Fig. \ref{fig9}.

In this simple wave packet propagation, we only considered axial
propagation along $z$. A discussion of paraxial effects and
diffraction losses are given later in the next subsection.
This one dimensional approach is shown to be valid in several
recent slow pulse experiments where the incident Gaussian beam
radius is much larger than the pinhole radius and the effects
of the far radial tail become negligible. We find that upon entering
the medium, the probe pulse rapidly slows down due to the sharply increased
refractive index of the BEC component. It then propagates with a
minimum speed around the center of the cloud, finally accelerates back
to its initial vacuum speed as it exits near the edge
of the cloud. According to this physical
picture, the average delay which is used in determining the
average group velocity is then lower than the optimal minimum
group velocity at the trap center. Indeed, from the figure we see
that the pulse arrives at the exit edge at about $0.6$ of the
predicted delay time estimated from a uniform density. In Ref.
\cite{morigi}, a group velocity of $9$ (m/s) was estimated for an
interacting gas of effective uniform density with $N=10^6$ atoms,
using the peak density from Thomas-Fermi approximation. When
taking into account the condensate edge effects by directly
propagating the wave equation in the scaled form, we obtained an
improved estimation of $15$ (m/s), which agrees quite well with
the reported value of $17$ (m/s). The average group velocity can be
made very close to the desired minimum speed by choosing a
density profile with sharp edges so that propagating light
traverses the medium with the minimum speed for a longer fraction
of total time spent in the medium. Ideally, a step function type density
distribution along the axial direction would achieve this.

\begin{figure}[t]
\centerline{\epsfig{file=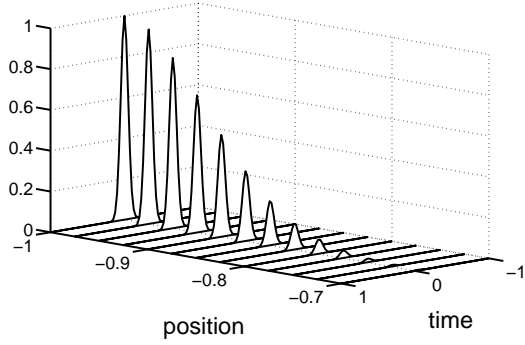,width=7.cm}\\[12pt]}
\caption{Absorption of an off-resonant pulse with detuning
$\Delta=3\gamma$, $\gamma=(2\pi)10.01$ (MHz), and $T=43$ (nK). All
other parameters are the same as in Fig. \ref{fig6}.Position and
time are given in dimensionless units explained in the text.}
\label{fig10}
\end{figure}

To complete this section, we also briefly studied absorption of
the pulse. We examined the absorption length in detail using the
propagation equation Eq.(\ref{seq}). At a large detuning of
$\Delta=3\gamma$ we observe a penetration into only the first
$15\%$ of the cloud as demonstrated in Fig. \ref{fig10}. By
controlling the absorption length via detuning, interaction of
light with only a selected part of the condensate might be
realized in such a simple way.

\subsection{Transverse diffraction effects}

\begin{figure}
\centerline{\epsfig{file=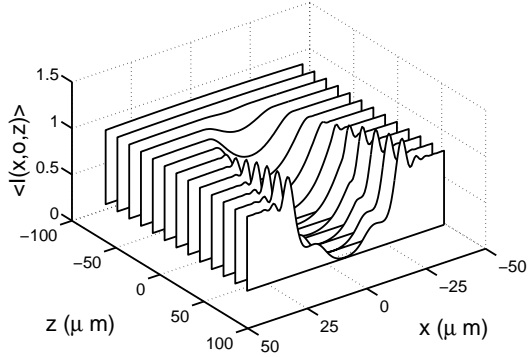,width=7.cm}\\[12pt]}
\caption{Diffraction pattern of a beam of radius 0.5 (mm) as it
propagates through a Bose condensate at $T=0.3T_C$. All other
parameters are the same as in Fig. \ref{fig6}.}
\label{difpat2.eps}
\end{figure}

The discussion of radial density profile effects requires the
inclusion of propagational diffraction. In practice, one can
always choose appropriate trap parameters and suitably
focused incident lasers to avoid most diffractive losses \cite{hau}.
To examine the diffraction of a probe beam as it propagates
through the condensate, we now include the previously neglected
term $\nabla^2_\bot{\cal E}$ to the wave equation. Employing
a standard pseudo-spectral split operator method for wave packet
propagation \cite{garraway} we have simulated the propagation of
a pulse of radial cross section size $0.5$ (mm) in accordance
with the experiment of Hau {\it et al.} \cite{hau}.
Figure \ref{difpat2.eps} illustrates its axial evolution of
the time averaged intensity pattern.
In general, we find essentially one dimensional propagation
equation Eq. (\ref{seq}) adequately describes near axis propagation
as performed in the experiment \cite{hau,inouye}. In the Fig.
\ref{difpat2.eps}, we only plotted the $y=0$ plane because of
its axial symmetry. Within the pinhole dimension set at $\sim
15\mu$m by experimental arrangement, and for detection close to
the atomic cloud, diffraction effects are seen to be
qualitatively negligible. Thus the large incident wave radius
effectively assures a plane wave signal for the cloud size which
is much smaller. The only significant effect of the transverse
diffraction is then the diffraction loss of the pulse amplitude.
In the experiment by Hau {\it et al.} \cite{hau}, it was
estimated that $80\%$ transmission of the probe pulse should occur
under the EIT condition, yet even at high temperatures the
observed transmission was limited to about $30\%$ \cite{hau}.
Since Fig. \ref{difpat2.eps} indicates a lower output intensity
than found from the one dimensional axial propagation model, we
suspect that the additional reduction in the pulse amplitude can
be due to the transverse diffraction loss.

\section{Conclusion}
We have performed a detailed investigation of Raman-matched
propagation of a probe laser pulse through a BEC.
We have discussed of the interpretations and
definitions of the group velocity, and argued that a complex valued
group velocity should be used when strong amplitude-phase
correlation exists. We have adopted the highly successful
two component model for atomic condensate density profile
and compared the differences of slow light propagation
through an interacting and an ideal gas BEC. Based on the
broadened condensate profile of an repulsively interacting
Na atom BEC, we have obtained group velocity values of the
same order of magnitude as observed in the experiment \cite{hau}.
Furthermore our simulations for this experiment explains
the qualitative behavior observed in the temperature dependence
of the group velocity at low temperatures \cite{hau}. In the
ultracold regime deep below the BEC transition temperature,
including the atom-atom interaction and the inhomogeneous
density profile results in
two orders of magnitude improvement in the
quantitative description of the group velocity behavior \cite{morigi}.
We also observed an increase in
the group velocity at larger atomic scattering length.

When comparing the detailed spatial averaging of operationally
defined delay time, we find that a sharp edged density profile yields
lower average group velocity $\langle v_g\rangle$, i.e. closer to
its local minimum $v_g(0)$, typically obtained based on the estimation
of group velocity using the peak density.
Our detailed numerical simulations confirm $\langle v_g\rangle\approx
v_g(0)/0.6$, which gives essentially the same $\langle v_g\rangle$ value
as reported \cite{hau}.

We have also compared the One dimensional numerical simulations
with a paraxial approximation including the diffraction effect
during the probe pulse propagation. We find that under the
experimental conditions, the use of a small pinhole to select
probe light rays close to the x-axis is a reasonable approach as
the diffractive effects are negligible apart from a total
loss of the overall signal level. Thus, our studies also justify
the one dimensional approach normally used in EIT studies.

\section{acknowledgments}
During the course of this work, one of us (L.Y.) was a participant
of the recent workshop on quantum degenerate gases at the Lorentz
Center, University of Leiden. L.Y. thanks H. Stoof for the
hospitality. This work is supported by the NSF grant No.
PHY-9722410.

%--------------------------------------------------------------------
%\begin{references}

%\end{references}
%-------------------------------------------------------------------------

%--------------------------------------------------------------------------
\end{document}